\documentstyle[12pt]{article}
\def\ba{\begin{array}}
\def\ea{\end{array}}
\def\be{\begin{equation}\begin{array}{l}}
\def\ee{\end{array}\end{equation}}
\def\bea{\begin{equation}\begin{array}{l}}
\def\eea{\end{array}\end{equation}}
\def\f{\frac}

\def\la{\lambda}
\def\vv{\f{V}{\la^d}}
\def\si{\Sigma}
\def\bi{\bibitem}
\def\c{\cite}

\title{ {\bf On Schwarzschild Black Hole in Large N Matrix theory} }
\author{ Yi-Hong Gao and Wei Zhang \thanks{zhwei@itp.ac.cn} \\
 Institute of Theoretical Physics , Academia Sinica,\\
 P.O.Box 2735, Beijing, 100080, China  }
\date{ }
\begin{document}
\maketitle

\vskip 0.8in

\begin{center}{\large \bf Abstract} \end{center}
\quad
D0-brane gas picture of Schwarzschild black hole (SBH) is considered in the large N regime of Matrix theory. An entropy formula, which reproduces the thermodynamics of SBH in the large N limit for all dimensions ($D\geq 6$), is proposed. The equations of states for low temperature are obtained. We also give a proof of the Newton gravitation law between two SBHs, whose masses are not equal. Our result in some extent supports that the physics of Matrix theory is Lorentz invariant.

\newpage

\baselineskip 18pt
\section{Introduction}
\quad
In recent years, we have got better understanding of the nonperturbative aspects of string theory with the help of the concept of D-brane \c{d} \c{pol1} \c{pol2}. An underlying M theory \c{tow} was proposed and 
a microscopic formulation of M theory--Matrix theory \c{bf}\c{se} was proposed. One of the important success of M theory is the derivation of Bekenstein-Hawing entropy formula by counting microscopic states \c{str}. Banks, Fischler, Klebanov and Susskind (BFKS)\c{ba1}\c{ba2} first explore Schwarzschild black hole (SBH) in matrix theory \c{kl} \c{ba3} \c{hor} \c{mi1} \c{ml} \c{ml2} \c{das}\c{low} \c{ram}. They suggested that  a SBH in Matrix theory can be viewed as Boltzmann gas of D0-branes. The D0-branes are distinguishable because they are tethyed to some classical background, which breaks the gauge symmetry (the source of the effective statistics of the D0-brane gas) completely. Minic further suggested that D0-brane gas should obey infinite statistics \c{mi1}. The approaches of BFKS \c{ba1} \c{ba2}  are based on the technical assumption of $ N\sim S $. That is their result are based on the Lorentz invariance of Matrix theory, which has not been proved yet.

In this paper, we propose an entropy formula in matrix theory to explore the SBH in the large N limit in detail. In our approach, the results of D0-brane gas fit those of SBH in all dimensions ($D\geq 6$). At the same time, we can get the low temperature equation of state, which was proposed by BFKS without proof. We will also derive the long distance Newton gravitation law for two SBHs, whose mass are not necessarily equal. This is the generalization of the result of BFKS. \\

\section{Schwarzschild black hole in large N limit}

We consider the system in D-dimension spacetime with 11-D dimension compactified. The transverse dimension of uncompactified space is d=11-D-2. For simplicity we consider Tordial compactification and assume that the torus is square with equal circumference given by L. We view the Schwarzschild black hole (with radius $R_s$ in rest frame) in DLQC version of matrix theory with longitudinal momentum $P_-=\f{N}{R}$ (where R is the radius of compactified circle in the longitudinal dimension) as D0-brane gas of number N, which obey infinite statistics \c{st}\c{vo}\c{mi1}. 

For infinite statistics gas\c{to}, one has
\bea
F=-NT ln \vv\\
S=\f{d}{2}N [1+ln(\vv)^{\f{2}{d}}]\\
\mu=F/N\\
\eea
where $V=R^d_s$ is the volume of D0-brane gas in transverse dimension, and $\la = (\f {2 \pi}{mT})^{1/2}$ is the thermodynamical wave length. There exists a critical temperature $T_c=\f{2 \pi}{mV^{\f {2}{d}}} \sim \f{R}{R^2_s}$ in this  system. At the critical temperature, $\mu=0$ and the average particle number divergents \c{to}. Let us consider the case near the critical temperature. We expand the right hand of (1) about the small parameter $t=(\f{V}{\la^d})^{\f{2}{d}}-1$,
\bea
S=\f{d}{2} N [1+ln(1+t)]\\
\sim \f{d}{2}N (\vv)^{\f{2}{d}}\\
\eea
\bea
S\sim NTR^2_s/R\\
=N(\f{T}{T_c})
\eea
We will ignore the numerical coefficient of order 1 through this paper.

It is easy to see that $S \sim N $, when $T=T_c$. This is just the case considered by BFKS and Minic. One can see from (1) that $S \rightarrow -\infty$, when $T \rightarrow 0$. This violates the third law of thermodynamics. It seems that infinite statistics in not suitable for the case of $N>>S (T<<T_c)$.  However formula (3) is consistent with the third thermodynamical law. We propose that formula (3) is the correct entropy formula for D0-brane gas, which describes SBH. We would like to give more evidence for the valid of the entropy formula (3).

Formula (3) is consistent with the estimate of Li's \c{ml}. One uses the Viral theory for the D0-brane gas.
\bea
N \f{1}{2} m<v^2> \sim E \sim TS\\
v \sim R_s T
\eea
where $m=\f{1}{R}, R_s$ is the typical distance and T is typical frequency of the system. From (4) we have the following entropy formula
\be
S=NTR^2_s/R
\ee

Equation (3) is also consistent with the cluster picture of D0-brane gas proposed by Li and Martinec \c{ml2}
\be
E=(\f{S}{R_s})^2 \f{R}{N}=ST
\ee

For Schwarzschild black hole at rest frame $T_H \sim 1/R_s$. In the boosted frame (with longitudinal momentum $P_-=\f{N}{R}$), 
\be
T_H=\f{MR}{NR_s}
\ee
On the other side, applying our entropy formula (3), we can get
\be
E=TS=TNTR^2_S/R=M^2R/N
\ee
That is $T=T_H$.

We calculate the mass of the black hole as
\bea
M^2=\f{N}{R}E(N,S)\\
=\f{N}{R}ST\\
=\f{N}{R}S\f{SR}{NR^2_s}\\
=\f{S^2}{R^2_s}
\eea
M calculated in this way is independent of N as it should be and the relation $S=MR_s$ is true for all dimension. Combining the above result and the fact that \c{ml} the dominate interaction among D0-branes at low temperature is of the form $G_D(\f{N}{R})^2\f{v^3}{r^{d-1}}$, one find that
\bea
M=R^{d-1}_s/G_D\\
S=R_s^d/G_D
\eea
which agrees with the usual thermodynamics of black hole.

\section{State of equation of supersymmetric Yang-Mill theory}
\quad
In this section we will derive the state of equation of supersymmetric Yang-Mill theory. We consider the general case of k=11-D dimensions are compactified. First we can get D0-brane from Dk-brane through T-duality. Since the duality is done in the transverse dimension, the temperature of the wrapped k-branes should be the same as that of D0-branes. From string duality in matrix model \c{su1} \c{ga}, one has
\be
\si=\f{l^3_{11}}{RL}
\ee 
The Newton constant in D-k dimensional spacetime is $G_{D-k}=\f{l^9_{11}}{L^k}$
and the coupling constant is $g^2=\f{l^{3k-6}_{11}R^{3-k}}{L^{k}}$. 
Using our formula (3) for entropy and (10), we have
\be
S=(NT)^{\f{D-2}{D-4}}G^{\f{2}{7-k}}_DR^{-\f{9-k}{7-k}}
\ee 
The above formula can be rewritten as
\be
S=(NT)^{\f{D-2}{D-4}}\si^{\f{(5-k)k}{7-k}}(g^2)^{\f{k-3}{7-k}}
\ee
For k=3, D=8
\be
S=(NT\si)^{\f{2}{3}}
\ee
the result agrees with that proposed by BFKS \c{ba1}. For K=1, D=10,
\be
S=(NT)^{\f{4}{3}}g^{-\f{2}{3}}\si^{\f{2}{3}}
\ee
which is the result proposed in \c{kl}.
If the condition $N\sim S$ is used, 
\be
S=N^{\f{1}{2}}E^{\la}\si^{k(1-\la)}g^a
\ee
where $2\la=\f{D-2}{D-4}, g=\f{8-D}{D-4}$. This is the result of k for all dimension for the case $N\sim S$ \c{kl}.

There are many evidences showing that there exists a critical temperature at which some kind of phase transition occurs. It is show that $T_{crit}=\f{1}{(N\si^3)^{1/3}}$ under the assumption $N\sim S$\c{ba1}.
From
\bea
N=S=R^6_s/G_8\\
\si=\f{G^{1/3}_8}{R}
\eea
It is easy to see that
\bea
T_{crit}=\f{R}{R^2_s}\\
=T_c
\eea 

\section{Interaction between black holes}
\quad
The interaction between black holes was considered by BFKS \c{ba2} for the equal mass case and their proof depends on the assumption $N\sim S$. Here we give a proof similar in spirit to that of BFKS. However our proof is valid for the general case and is independent of particular choice of N.
Consider two black hole with mass $M_1,M_2$ in restframe. The potential energy between them is V(r), where r is the the distance between the black holes. The corresponding configuration in matrix theory is two D0-brane bound states with distance r in transverse dimension. The longitudinal momentum is $P_-=\f{N}{R}$. Thus the total number of D-particle is N. Suppose the number of D-particles of the two bound states are $N_1,N_2$ respectively. From
\be
N_1+N_2=N
\ee
and the energy relation 
\be
\f{(M_1+M_2)^2R}{N}=\f{M^2_1R}{N_1}+\f{M^2_2R}{N_2}
\ee
we can determine $N_1, N_2$ as
\be
N_1=\f{M_1}{M_1+M_2} N,N_2=\f{M_2}{M_1+M_2}N
\ee
The fact
\be
E_i=\f{M^2_iR}{N_i}=\f{1}{2} \f{1}{R} v_i^2
\ee
(i=1,2) leads to
\be
v_1=v_2=\f{(M_1+M_2)R}{N}
\ee
The potential energy between the two bound states is
\bea
\f{(M_1+M_2+V(r))^2R}{N}-\f{(M_1+M_2)^2R}{N}\\
\sim \f{(M_1+M_2)R}{N}V(r)
\eea
which is equal to potential in the Hamiltonian of D0-branes
\be
PE=G\f{N_1N_2|v_1-v_2|^3}{R^2r^{d-1}}
\ee
Comparison of the two expression of the potential (24) and (25) leads to 
\be
V(r)=G\f{M_1M_2}{r^{d-1}} 
\ee

Thus we get the Newton potential between two black holes in the rest frame which is independent of R as expected.
Here we have used the potential $PE=G\f{N_1N_2|v_1-v_2|^3}{R^2r^{d-1}}
$ instead of $PE'=G\f{N_1N_2|v_1-v_2|^4}{R^2r^{d-2}}$, because the potential PE is more important at low temperature from the argument of Li's \c{ml}. What would happen if we use potential
\be
PE'=G\f{N_1N_2|v_1-v_2|^4}{R^3r^{d-2}}
\ee
instead of (25).
We will have
\be
V(r)=G\f{M_1M_2}{r^{d-2}} \f{M_1+M_2}{N}
\ee
If we use the condition 
\be
\f{(M_1+M_2)R}{N} r=R
\ee
equation (28)becomes (26).
Condition (29) is something like
\be
\f{MR}{N}R_s=R
\ee 
which is equivalent to $N \sim S$ or $T=T_c$.
The above result supports Li's argument on the fact that interaction PE is more important than interaction PE'at low temperature.\\

{ \large \bf Acknowledgment}\\

We would like to thank Han-Yin Guo, Yun-kau Lau, L.Miao, Hua-ling Shi, and Chi Xiong for helpful discussions.\\

\newpage

\end{document}